\documentclass[runningheads]{llncs}
\title{A High-Order Discontinuous Galerkin Solver with Dynamic Adaptive Mesh Refinement to Simulate Cloud Formation Processes }
\titlerunning{A High-Order DG Solver to Simulate Cloud Formation Processes}
\author{Lukas Krenz\orcidID{0000-0001-6378-0778} \and Leonhard Rannabauer \and Michael Bader}
\authorrunning{L.\ Krenz, L.\ Rannabauer, M.\ Bader}
\institute{Department of Informatics, Technical University of Munich\\
  \email{lukas.krenz@in.tum.de}, \email{rannabau@in.tum.de}, \email{bader@in.tum.de}
} 
\usepackage[utf8]{inputenc}
\usepackage[american]{babel}
\usepackage[autostyle, english = american]{csquotes}
\usepackage{cite}
\newcommand{\citeauthor}{???}

\usepackage{caption}
\usepackage{subcaption}
\usepackage{xparse} 
\usepackage{etoolbox} 
\usepackage{xstring} 
\usepackage{xpatch}
\usepackage{xcolor}
\usepackage{amsmath}
\usepackage{amsfonts}
\usepackage{amssymb}
\usepackage{mathtools} 
\usepackage{nicefrac}
\usepackage{physics} 
\usepackage{varioref}
\usepackage{nicefrac}
\usepackage{physics} 
\usepackage[separate-uncertainty]{siunitx}
\usepackage{hyperref}
\usepackage[capitalise]{cleveref}
\usepackage{graphicx}
\usepackage{multimedia}
\graphicspath{{.}}
\crefformat{equation}{\eqA{}Eq.~\eqB #2#1#3)}
\crefmultiformat{equation}{Eqs.~\eqMultiA#2#1#3\eqMultiB}%
{ and \eqMultiA#2#1#3\eqMultiB}{, \eqMultiA#2#1#3\eqMultiB}{ and~\eqMultiA#2#1#3\eqMultiB}
\newcommand{\eqA}{}
\newcommand{\eqB}{(}
\newcommand{\eqMultiA}{(}
\newcommand{\eqMultiB}{)}
\DeclareRobustCommand{\pcrefSingle}[1]{%
\begingroup%
  \renewcommand{\eqA}{(}\renewcommand{\eqB}{}%
\cref{#1}%
\endgroup%
}
\DeclareRobustCommand{\pcrefMulti}[1]{%
\begingroup%
    \renewcommand{\eqMultiA}{}\renewcommand{\eqMultiB}{}%
    (\cref{#1})%
\endgroup%
}
\DeclareRobustCommand{\pcref}[1]{%
\IfSubStr{#1}{,}{\pcrefMulti{#1}}{\pcrefSingle{#1}}%
}

\usepackage{bm}
\newcommand{\muscl}{\textsc{muscl}-Hancock}
\newcommand{\dg}{\textsc{dg}}

\newcommand{\aderdg}{\textsc{ader-dg}}
\newcommand{\amr}{\textsc{amr}}
\newcommand{\pde}{\textsc{pde}}
\newcommand{\tbb}{\textsc{tbb}}
\newcommand{\mpi}{\textsc{mpi}}

\newcommand{\softwareName}[1]{#1}
\newcommand{\exahype}{\softwareName{ExaHyPE}}
\newcommand{\exahypeengine}{\softwareName{ExaHyPE Engine}}

\newcommand{\Q}{\bm{Q}}
\newcommand{\gradQ}{\gradient{\Q}}
\newcommand{\Qrho}{\rho}
\newcommand{\Qj}{\rho \bm{v}}
\newcommand{\Qv}{\bm{v}}
\newcommand{\QE}{\rho E}
\newcommand{\potT}{\theta}

\newcommand{\pertubationPotT}{\theta'}
\newcommand{\stressT}{\bm{\sigma}}
\newcommand{\pressure}{p}

\newcommand{\cell}[1][]{C_{#1}}

\newcommand{\flux}{\bm{F}}
\newcommand{\viscFlux}{\flux^{v}}
\newcommand{\hyperFlux}{\flux^{h}}
\newcommand{\source}[1][]{
  \notblank{#1}{
S_{#1}
}{
\bm{S}
}
}
\newcommand{\intdcell}[1]{\int_{\cell} #1 \dd{\bm{x}}}
\newcommand{\tv}{\operatorname{TV}}

\begin{document}
\maketitle 
\begin{abstract}
We present a high-order discontinuous Galerkin (\dg{}) solver of the compressible Navier-Stokes equations for cloud formation processes.
The scheme exploits an underlying parallelized implementation of the \aderdg{} method with dynamic adaptive mesh refinement. 
We improve our method by a \pde{}-independent general refinement criterion, based on the local total variation of the numerical solution.
While established methods use numerics tailored towards the specific simulation, our scheme works scenario independent.
Our generic scheme shows competitive results for both classical \textsc{cfd} and stratified scenarios.
We focus on two dimensional simulations of two bubble convection scenarios over a background atmosphere.
The largest simulation here uses order 6 and 6561 cells which were reduced to 1953 cells by our refinement criterion.
\end{abstract}
\section{Introduction}
In this paper we address the resolution of basic cloud formation processes on modern super computer systems.
The simulation of cloud formations, as part of convective processes, is expected to play an important role in future numerical weather prediction~\cite{bauer2015quiet}.
This requires both suitable physical models and effective computational realizations. 
Here we focus on the simulation of simple benchmark scenarios~\cite{giraldo2008study}.
They contain relatively small scale effects which are well approximated with the compressible Navier-Stokes equations.
We use the \aderdg{} method of~\cite{dumbser2008unified}, which allows us to simulate the Navier-Stokes equations with a space-time-discretization of arbitrary high order.
In contrast to Runge-Kutta time integrators or semi-implicit methods, an increase of the order of \aderdg{} only results in larger computational kernels and does not affect the complexity of the scheme.
Additionally, \aderdg{} is a communication avoiding scheme and reduces the overhead on larger scale.
We see our scheme in the regime of already established methods for cloud simulations, as seen for example in~\cite{giraldo2008study,muller2010adaptive,muller2018strong}.

Due to the viscous components of the Navier-Stokes equations, it is not straightforward to apply the \aderdg{} formalism of~\cite{dumbser2008unified}, which addresses hyperbolic systems of partial differentials equations (\pde{}s) in first-order formulation.
To include viscosity, we use the numerical flux for the compressible Navier-Stokes equations of Gassner et al.~\cite{gassner2008discontinuous}.
This flux has already been applied to the \aderdg{} method in~\cite{dumbser2010arbitrary}.
In contrast to this paper, we focus on the simulation of complex flows with a gravitational source term and a realistic background atmosphere.
Additionally, we use adaptive mesh refinement (\textsc{amr}) to increase the spatial resolution in areas of interest.
This has been shown to work well for the simulation of cloud dynamics~\cite{muller2010adaptive}.
Regarding the issue of limiting in high-order \dg{} methods, we note that viscosity not only models the correct physics of the problem but also smooths oscillations and discontinuities, thus stabilizing the simulation.

We base our work on the \exahypeengine{} (\url{www.exahype.eu}), which is a framework that can solve arbitrary hyperbolic \pde\ systems.
A user of the engine is provided with a simple code interface which mirrors the parts required to formulate a well-posed Cauchy problem for a system of hyperbolic \pde{}s of first order.
The underlying \aderdg{} method, parallelization techniques and dynamic adaptive mesh refinement are available for simulations while the implementations are left as a black box to the user.
An introduction to the communication-avoiding implementation of the whole numerical scheme can be found in~\cite{charrier2018stop}.

To summarize, we make the following contributions in this paper:%
\begin{itemize}%
\item We extend the \exahypeengine{} to allow viscous terms.
\item We thus provide an implementation of the compressible Navier-Stokes equations.
  In addition, we tailor the equation set to stratified flows with gravitational source term.
  We emphasize that we use a standard formulation of the Navier-Stokes equations as seen in the field of computational fluid mechanics and only use small modifications of the governing equations, in contrast to a equation set that is tailored exactly to the application area.
\item We present a general \textsc{amr}-criterion that is based on the detection of outlier cells w.r.t.\ their total variation.
  Furthermore, we show how to utilize this criterion for stratified flows.
\item We evaluate our implementation with standard \textsc{cfd} scenarios and atmospheric flows and inspect the effectiveness of our proposed \amr{}-criterion.
  We thus inspect, whether our proposed general implementation can achieve results that are competitive with the state-of-the-art models that rely on heavily specified equations and numerics.
\end{itemize}
\section{Equation Set}
\newcommand{\diffCoeff}{\varepsilon}%
\newcommand{\hyperFluxDef}{
  \begin{pmatrix}
    \Qj \\
    \Qv  \otimes \Qj + \bm{I} \pressure  \\
    \Qv \cdot (\bm{I} \QE + \bm{I} \pressure)
  \end{pmatrix}
}%
\newcommand{\viscFluxDef}{
  \begin{pmatrix}
    0\\
     \stressT (\Q, \gradQ)  \\
     \Qv \cdot \stressT (\Q, \gradQ) - \kappa \gradient{T}
   \end{pmatrix}
}%
The compressible Navier-Stokes equations in the conservative form are given as
\begin{equation}
 \label{eq:equation-set} 
\quad
  \pdv{}{t}
  \underbrace{
  \begin{pmatrix}
    \Qrho\\
    \Qj\\
    \QE
    \end{pmatrix}}_{\Q}
  +
  \divergence{
  \underbrace{
  \left(
   \underbrace{\hyperFluxDef}_{\hyperFlux(\Q)}
+
\underbrace{\viscFluxDef}_{\viscFlux(\Q, \gradQ)}
  \right)}_{\flux(\Q, \gradQ)}}
 =
  \underbrace{
  \begin{pmatrix}
    \source[\Qrho\phantom{\Qrho}]\\
    \source[\Qj]\\
    \source[\QE]
    \end{pmatrix}}_{\source(\Q, \bm{x}, t)}
\end{equation}
with the vector of conserved quantities $\Q$, flux $\flux(\Q, \gradQ)$ and source $\source(\Q)$.
Note that the flux can be split into a hyperbolic part $\hyperFlux(\Q)$,
which is identical to the flux of the Euler equations,
and a viscous part $\viscFlux(\Q, \gradQ)$.
The conserved quantities
\(\Q\)
are the density $\Qrho$, the two or three-dimensional momentum $\Qj$ and the energy density $\QE$.
The rows of \cref{eq:equation-set} are the conservation of mass, the conservation of momentum and the conservation of energy.

The pressure $\pressure$ is given by the equation of state of an ideal gas
\begin{equation}
  \label{eq:eos}
  \pressure = (\gamma - 1) \left(\QE - \frac{1}{2} \left(\Qv \cdot \Qj \right) - gz \right).
\end{equation}
The term $gz$ is the geopotential height with the gravity of Earth $g$~\cite{giraldo2008study}.
The temperature $T$ relates to the pressure by the thermal equation of state
\begin{equation}
  \label{eq:temperature}
  \pressure = \Qrho R T,
\end{equation}
where $R$ is the specific gas constant of a fluid.

We model the diffusivity by the stress tensor
\begin{equation}
  \label{eq:stress-tensor}
  \stressT(\Q, \gradQ) =
  \mu
  \bigl(
  \left(\nicefrac{2}{3} \divergence{\Qv} \right) -
  \left( \gradient{\Qv} + \gradient{\Qv}^\intercal \right)
  \bigr),
\end{equation}
with constant viscosity $\mu$.
The heat diffusion is governed by the coefficient
\begin{equation}
  \label{eq:heat-conduction-coeff}
  \kappa = \frac{\mu \gamma}{\Pr} \frac{1}{\gamma - 1} R = \frac{\mu c_p}{\Pr},
\end{equation}
where the ratio of specific heats $\gamma$, the heat capacity at constant pressure $c_p$ and the Prandtl number $\Pr$ depend on the fluid.

Many realistic atmospheric flows can be described by a perturbation over a background state that is in hydrostatic equilibrium
\newcommand{\backgroundPressure}{\overline{\pressure}}
\newcommand{\backgroundRho}{\overline{\Qrho}}
\begin{equation}
  \label{eq:hydrostatic-balance}
  \pdv{}{z} \backgroundPressure{\left (z \right )} = -g \backgroundRho(z),
\end{equation}
i.e.\ a state, where the pressure gradient is exactly in balance with the gravitational source term $\source[\Qj] = - \bm{k} \Qrho g$.
The vector $\bm{k}$ is the unit vector pointing in $z$-direction.
The momentum equation is dominated by the background flow in this case.
Because this can lead to numerical instabilities, problems of this kind are challenging and require some care.
To lessen the impact of this, we split the pressure $\pressure = \backgroundPressure + \pressure'$ into a sum of the background pressure $\backgroundPressure(z)$ and perturbation $\pressure'(\bm{x}, t)$.
We split the density $\Qrho = \backgroundRho + \Qrho'$ in the same manner and arrive at
\begin{equation}
  \label{eq:momentum-equation-split}
  \pdv{\Qj}{t}+ \divergence{ \left(
    \Qv \otimes \Qj + \bm{I} \pressure'
    \right)
  } + \viscFlux_\Qj
  =
  -g \bm{k} \Qrho'.
\end{equation}
Note that a similar and more complex splitting is performed in~\cite{muller2010adaptive,giraldo2008study}.
In contrast to this, we use the true compressible Navier-Stokes equations with minimal modifications.

\section{Numerics}
The \exahypeengine{} implements an \aderdg{}-scheme and a \muscl{} finite volume method.
Both can be considered as instances of the more general \textsc{PnPm} schemes of~\cite{dumbser2008unified}.
We use a Rusanov-style flux that is adapted to \pde{}s with viscous terms~\cite{gassner2008discontinuous,fambri2017space}.
The finite volume scheme is stabilized with the van Albada limiter~\cite{van1997comparative}.
The user can state dynamic \amr{} rules by supplying custom criteria that are evaluated point-wise.
Our criterion uses an element-local error estimate based on the total variation of the numerical solution.
We exploit the fact that the total variation of a numerical solution is a perfect indicator for edges of a wavefront.
Let $\bm{f}(\bm{x}): \mathbb{R}^{N_\text{vars}} \to \mathbb{R}$ be a sufficiently smooth function that maps the discrete solution at a point $\bm{x}$ to an arbitrary indicator variable.
The total variation (\textsc{tv}) of this function is defined by
\begin{equation}
  \label{eq:tv}
  \tv \left[ f(\bm{x}) \right] =
  \left\Vert
\intdcell{ \vert \gradient{f \left( \bm{x} \right)} \vert }
\right\Vert_1
\end{equation}
for each cell.
The operator $\Vert \cdot \Vert_1$ denotes the discrete $L_1$ norm in this equation.
We compute the integral efficiently with Gaussian quadrature over the collocated quadrature points.
\newcommand{\mean}{\mu}%
\newcommand{\std}{\sigma}%
\newcommand{\variance}{\std^2}%
\newcommand{\Trefine}{T_\text{refine}}%
\newcommand{\Tdelete}{T_\text{coarsen}}%
How can we decide whether a cell is important or not?
To resolve this conundrum, we compute the mean and the population standard deviation of the total variation of all cells.
It is important that we use the method of~\cite{chan1982updating} to compute the modes in a parallel and numerical stable manner.
A cell is then considered to contain significant information if its deviates from the mean more than a given threshold.
This criterion can be described formally by
\begin{equation}
  \label{eq:refinement-criterion}
  \operatorname{evaluate-refinement}(\Q, \mu, \sigma) =
  \begin{cases}
    \text{refine} & \text{if } \tv(\Q) \geq \mu + \Trefine \sigma, \\
    \text{coarsen} & \text{if } \tv(\Q) < \mu + \Tdelete \sigma, \\
    \text{keep} & \text{otherwise}.
    \end{cases}
\end{equation}
The parameters $\Trefine > \Tdelete$ can be chosen freely.
Chebyshev's inequality
\begin{equation}
  \label{eq:chebychev}
  \mathbb{P}\bigl(\vert X - \mu \vert \geq c \sigma \bigr) \leq \frac{1}{c^2},
\end{equation}
with probability $\mathbb{P}$ guarantees that we neither mark all cells for refinement nor for coarsening.
This inequality holds for arbitrary distributions under the weak assumption that they have a finite mean $\mu$ and a finite standard deviation $\sigma$~\cite{wasserman2004all}.
Note that subcells are coarsened only if all subcells belonging to the coarse cell are marked for coarsening.
In contrast to already published criteria which are either designed solely for the simulation of clouds~\cite{muller2010adaptive} or computationally expensive~\cite{fambri2017space}, our criterion works for arbitrary \pde{}s and yet, is easy to compute and intuitive.

\section{Results}\label{sec:results}
In this section, we evaluate the quality of the results of our numerical methods and the scalability of our implementation.
We use a mix of various benchmarking scenarios.
After investigating the numerical convergence rate, we look at three standard \textsc{cfd} scenarios: the Taylor-Green vortex, the three-dimensional Arnold-Beltrami-Childress flow and a lid-driven cavity flow.
Finally, we evaluate the performance for stratified flow scenarios in both two and three dimensions.

\subsection{CFD Testing Scenarios}
\begin{figure}[tb]
  \centering
  \includegraphics{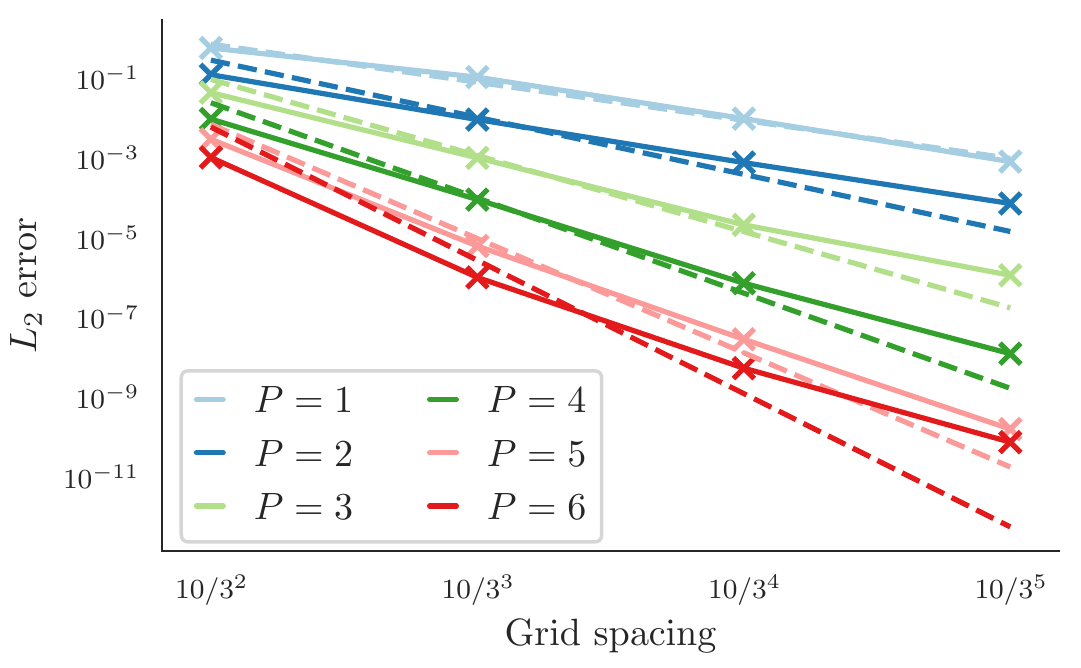}
  \caption{\label{fig:convergence-test}Mesh size vs.\ error for various polynomial orders $P$.
    Dashed lines show the theoretical convergence order of $P+1$.}
\end{figure}

We begin with a manufactured solution scenario which we can use for a convergence test.
We use the following constants of fluids for all scenarios in this section:
\begin{equation}
  \gamma = 1.4, \quad \Pr = 0.7, \quad c_v = 1.0.
\end{equation}
Our description of the manufactured solution follows~\cite{dumbser2010arbitrary}.
To construct this solution, we assume that
\begin{align}\label{eq:manufactured-solution}
  \begin{split}
    \pressure(\bm{x}, t) &= \nicefrac{1}{10} \cos( \bm{k} \bm{x} - 2 \pi t ) + \nicefrac{1}{\gamma}, \\
    \Qrho(\bm{x}, t) &= \nicefrac{1}{2} \sin (\bm{k} \bm{x} - 2 \pi t) + 1, \\
    \Qv(\bm{x}, t) &= \bm{v_0} \sin(\bm{k} \bm{x} - 2 \pi t),
  \end{split}
  \end{align}
solves our \pde{}.
We use the constants $\bm{v_0} = \nicefrac{1}{4} \left( 1, 1 \right)^\intercal$, $\bm{k} = \nicefrac{\pi}{5} \left( 1, 1 \right)^\intercal$ and simulate a domain of size $\left[ 10 \times 10 \right]$ for $\SI{0.5}{\s}$.
The viscosity is set to $\mu = 0.1$.
Note that \Cref{eq:manufactured-solution} does not solve the compressible Navier-Stokes equations \cref{eq:equation-set} directly.
It rather solves our equation set with an added source term which can be derived with a computer algebra system.
We ran this for a a combination of orders $1, \ldots, 6$ and multiple grid sizes.
Note that by order we mean the polynomial order throughout the entire paper and not the theoretical convergence order.
For this scenario, we achieve high-order convergence (\cref{fig:convergence-test}) but notice some diminishing returns for large orders.

After we have established that the implementation of our numerical method converges, we are going to investigate three established testing scenarios from the field of computational fluid mechanics.
A simple scenario is the Taylor-Green vortex.
Assuming an \textit{incompressible} fluid, it can be written as
\begin{align}
  \label{eq:taylor-green}
  \begin{split}
  \Qrho(\bm{x}, t) &= 1,\\
  \Qv(\bm{x}, t) &= \exp(-2 \mu t)
  \begin{pmatrix}
    \phantom{-}\sin(x) \cos(y) \\
- \cos(x) \sin(y) 
    \end{pmatrix}, \\
  \pressure(\bm{x}, t) &= \exp(-4 \mu t) \, \nicefrac{1}{4} \left( \cos(2x) + \cos(2y) \right) + C.
  \end{split}
\end{align}
The constant $C = \nicefrac{100}{\gamma}$ governs the speed of sound and thus the Mach number $\text{Ma} = 0.1$~\cite{dumbser2016high}.
The viscosity is set to $\mu = 0.1$. 

\begin{figure}[tb]
  \centering
  \begin{subfigure}[t]{0.473\textwidth}
    \centering
    \includegraphics{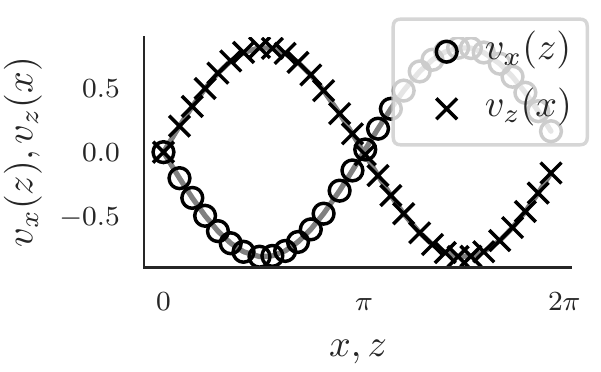}
    \caption{\label{fig:taylor-green}Our result (markers) of the Taylor-Green vortex vs.\ the analytical solution (lines) \cref{eq:taylor-green}.
    The plot shows two velocity slices, the respective other coordinate is held constant at a value of $\pi$.}
  \end{subfigure}\qquad%
\begin{subfigure}[t]{.473\textwidth}
  \centering
    \includegraphics{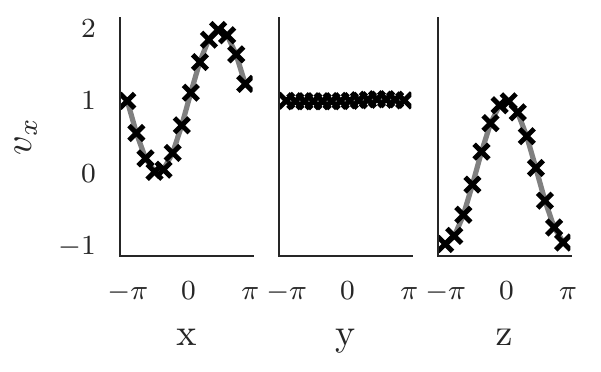}
    \captionof{figure}{\label{fig:abc-flow}Our approximation (markers) of the \textsc{abc}-flow vs.\ analytical solution (lines, \cref{eq:abc-flow}).
    All other axes are held constant at a value of 0. We show every 6th value.}
\end{subfigure}
  \caption{\label{fig:cdf-results}Two-dimensional \textsc{cfd} scenarios}
\end{figure}

We simulate on the domain $[0,2\pi]^2$ and impose the analytical solution at the boundary.
A comparison at time $t = 10.0$ of the analytical solution for the pressure with our approximation (\cref{fig:taylor-green}) shows excellent agreement.
Note that we only show a qualitative analysis because this is not an exact solution for our equation set as we assume compressibility of the fluid.
This is nevertheless a valid comparison because for very low Mach numbers, both incompressible and compressible equations behave in a very similarly.
We used an \aderdg{}-scheme of order $5$ with a grid of $25^2$ cells.

The Arnold-Beltrami-Childress (\textsc{abc}) flow is similar to the Taylor-Green vortex but is an analytical solution for the three-dimensional \textit{incompressible} Navier-Stokes equations~\cite{tavelli2016staggered}.
It is defined in the domain \( \left[ -\pi, \pi \right]^3 \) as
\begin{align}
  \label{eq:abc-flow}
  \begin{split}
  \Qrho(\bm{x}, t) &= 1,\\
  \Qv(\bm{x}, t) &= \phantom{-} \exp(-1\mu t)
  \begin{pmatrix}
    \sin(z) + \cos(y)\\
    \sin(x) + \cos(z)\\
    \sin(y) + \cos(x)
  \end{pmatrix}, \\
  \pressure(\bm{x}, t) &= -\exp(-2 \mu t) \, \left(\cos(x)\sin(y) + \sin(x)\cos(z) + \sin(z)\cos(y)\right)
  + C.
  \end{split}
\end{align}
The constant $C = \nicefrac{100}{\gamma}$ is chosen as before.
We use a viscosity of $\mu = 0.01$ and analytical boundary conditions.
Our results (\cref{fig:abc-flow}) show a good agreement between the analytical solution and our approximation with an \aderdg{}-scheme of order $3$ with a mesh consisting of $27^3$ cells at time $t = \SI{0.1}{\s}$.
Again, we do not perform a quantitative analysis as the \textsc{abc}-flow only solves our equation set approximately.

\begin{figure}[tb]
\centering
  \begin{minipage}[t]{.473\textwidth}
    \centering
    \includegraphics{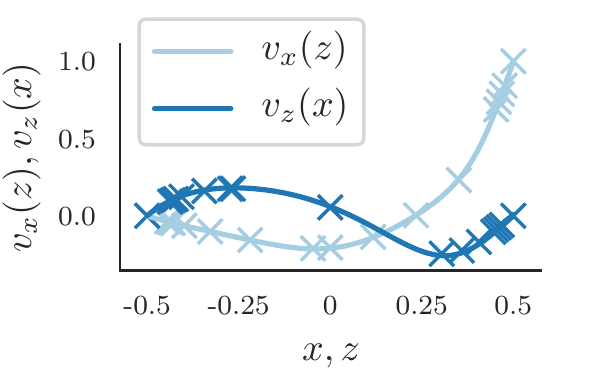}
    \caption{\label{fig:cavity-flow}Our approximation (solid lines) of the lid-driven cavity flow vs.\ reference solution (crosses) of~\cite{ghia1982high}.
    The respective other coordinate is held constant at a value of 0.}
  \end{minipage}\quad%
\begin{minipage}[t]{.473\textwidth}
  \centering
    \includegraphics[width=\textwidth]{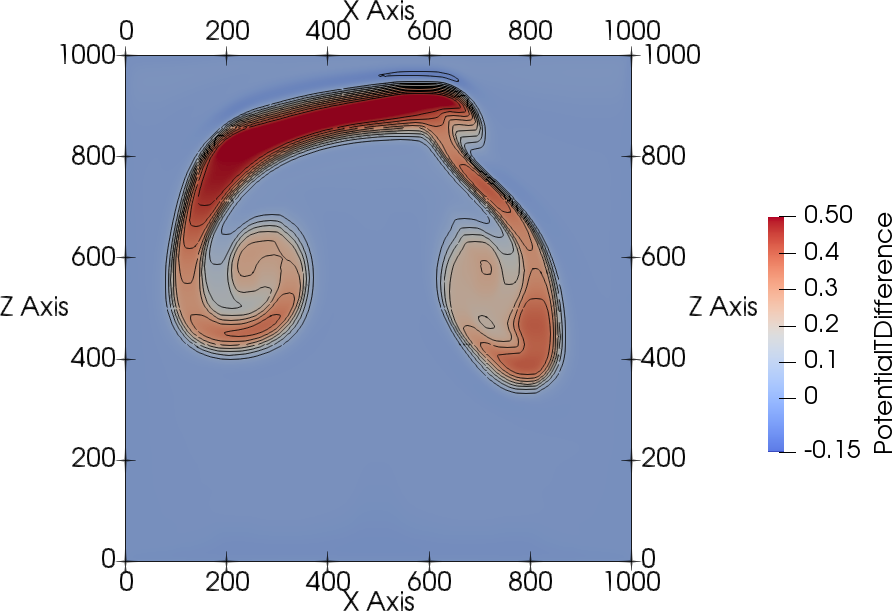}
    \captionof{figure}{\label{fig:two-bubbles-fv}%
    Colliding bubbles with \muscl{}. Contour values for potential temperature perturbation are $-0.05, 0.05, 0.1, \ldots 0.45$.}
\end{minipage}
\end{figure}

As a final example of standard flow scenarios, we consider the lid-driven cavity flow where the fluid is initially at rest, with $\Qrho = 1$ and $ \pressure(\bm{x}) = \nicefrac{100}{\gamma}$.
We consider a domain of size $\SI{1}{m} \times \SI{1}{\m}$ which is surrounded by no-slip walls.
The flow is driven entirely by the upper wall which has a velocity of $v_x = \SI{1}{\m/\s}$.
The simulation runs for $\SI{10}{\s}$.
Again, our results (\cref{fig:cavity-flow}) have an excellent agreement with the reference solution of~\cite{ghia1982high}.
We used an \aderdg{}-method of order $3$ with a mesh of size $27^2$.

\subsection{Stratified Flow Scenarios}
Our main focus is the simulation of stratified flow scenarios.
In the following, we present bubble convection scenarios in both two and three dimensions.
With the constants
\begin{equation}\label{eq:atmosphere-constants}
    \gamma = 1.4 ,\quad \Pr =  0.71 ,\quad R = 287.058 ,\quad p_0 = 10^5 \SI{}{\Pa}, \quad g = \SI{9.8}{m/s^2},
\end{equation}
all following scenarios are described in terms of the potential temperature
\begin{equation}
  \potT = T \left( \frac{p_0}{p} \right)^{R/c_p},
\end{equation}
with reference pressure $p_0$~\cite{muller2010adaptive,giraldo2008study}.
We compute the initial background density and pressure by inserting the assumption of a constant background energy in \cref{eq:hydrostatic-balance}.
The background atmosphere is then perturbed.
We set the density and energy at the boundary such that it corresponds to the background atmosphere.
Furthermore, to ensure that the atmosphere stays in hydrostatic balance, we need to impose the viscous heat flux
\begin{equation}
  \label{eq:atmosphere-bc}
  \viscFlux_{\QE} = \kappa \pdv{\overline{T}}{z}.
\end{equation}
at the boundary~\cite{giraldo2008study}.
In this equation, $\overline{T}(z)$ is the background temperature at position $z$, which can be computed from \cref{eq:hydrostatic-balance,eq:eos}.

Our first scenario is the colliding bubbles scenario~\cite{muller2010adaptive}.
We use perturbations of the form
\begin{equation}
  \label{eq:bubbles-pertubation}
  \pertubationPotT =
  \begin{cases}
    A & r \leq a, \\
    A \exp \left( - \frac{(r-a)^2}{s^2} \right) & r > a,
    \end{cases}
\end{equation}
where $s$ is the decay rate and $r$  is the radius to the center
\begin{equation}
  \label{eq:radius}
  r^2 = \Vert \bm{x} - \bm{x_c} \Vert_2,
\end{equation}
i.e., $r$ denotes the Euclidean distance between the spatial positions $\bm{x} = (x, z)$ and the center of a bubble $\bm{x_c} = (x_c, z_c)$ -- for three-dimensional scenarios $\bm{x}$ and $\bm{x_c}$ also contain a $y$ coordinate.

\begin{figure}[tb]
  \centering
  \includegraphics[width=1.0\textwidth]{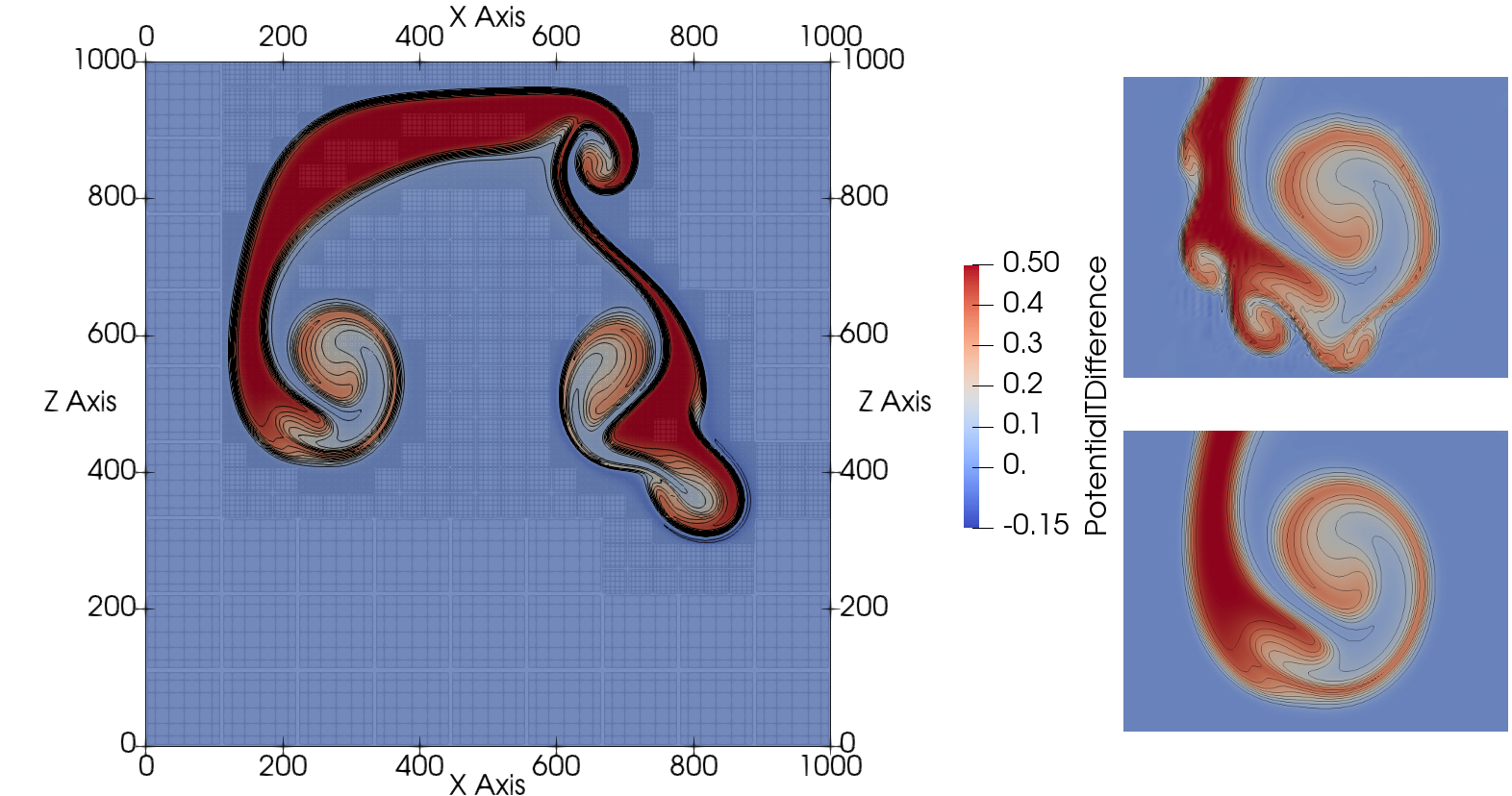}
  \caption{\label{fig:two-bubbles-ader}Left: Colliding Bubbles with \aderdg{}. Contour values for potential temperature perturbation are $-0.05, 0.05, 0.1, \ldots 0.45$.\\
  Right: Comparison of small scale structure between order 3 (top) and order 6 (bottom).}
\end{figure}
We have two bubbles, with constants
\begin{equation}
  \label{eq:bubbles-values}
\begin{alignedat}{6}
  & \text{warm:} \qquad && A = \SI{0.5}{\K}, \quad&& a = \SI{150}{\m}, \quad&& s = \SI{50}{\m}, \quad&& x_c = \SI{500}{\m,} \quad&& z_c = \SI{300}{\m},\\
  & \text{cold:} \qquad && A = \SI{-0.15}{\K}, \quad&& a = \SI{0}{\m}, \quad&& s = \SI{50}{\m}, \quad&& x_c = \SI{560}{\m}, \quad&& z_c = \SI{640}{\m}.
  \end{alignedat}
\end{equation}
Similar to~\cite{muller2010adaptive}, we use a constant viscosity of $\mu = 0.001$ to regularize the solution.
Note that we use a different implementation of viscosity than~\cite{muller2010adaptive}.
Hence, it is difficult to compare the parametrization directly.
We ran this scenario twice:
once without \amr{} and a mesh of size $\SI{1000/81}{\m} = \SI{12.35}{\m}$ and once with \amr{} with two adaptive refinement levels and parameters $\Trefine = 2.5$ and $\Tdelete = -0.5$.
For both settings we used polynomials of order 6.
We specialize the \amr{}-criterion~\pcref{eq:refinement-criterion} to our stratified flows by using the potential temperature.
This resulted in a mesh with cell-size lengths of approx.\ \SI{111.1}{\m}, \SI{37.04}{\m}, and \SI{12.34}{\m}.
The resulting mesh can be seen in \cref{fig:two-bubbles-ader}.

We observe that the $L_2$ difference between the potential temperature of the \amr{} run, which uses 1953 cells, and the one of the fully refined run with 6561 cells, is only $1.87$.
The relative error is \num{2.6e-6}.
We further emphasize that our \amr{}-criterion accurately tracks the position of the edges of the cloud instead of only its position.
This is the main advantage of our gradient-based method in contrast to methods working directly with the value of the solution, as for example~\cite{muller2010adaptive}.
Overall, our result for this benchmark shows an excellent agreement to the previous solutions of~\cite{muller2010adaptive}.
In addition, we ran a simulation with the same settings for a polynomial order of 3.
The lower resolution leads to spurious waves~(\cref{fig:two-bubbles-ader}) and does not capture the behavior of the cloud.

Furthermore, we simulated the same scenario with our \muscl{} method, using $7^2$ patches with $90^2$ finite volume cells each.
As we use limiting, we do not need any viscosity.
The results of this method (\cref{fig:two-bubbles-fv}) also agree with the reference but contain fewer details.
Note that the numerical dissipativity of the finite volume scheme has a smoothing effect that is similar to the smoothing caused by viscosity.

\begin{figure}[tb]
  \centering
  \begin{subfigure}[t]{0.5\textwidth}
    \centering
    \includegraphics[scale=0.16, trim={0cm 0 32cm 0},clip]{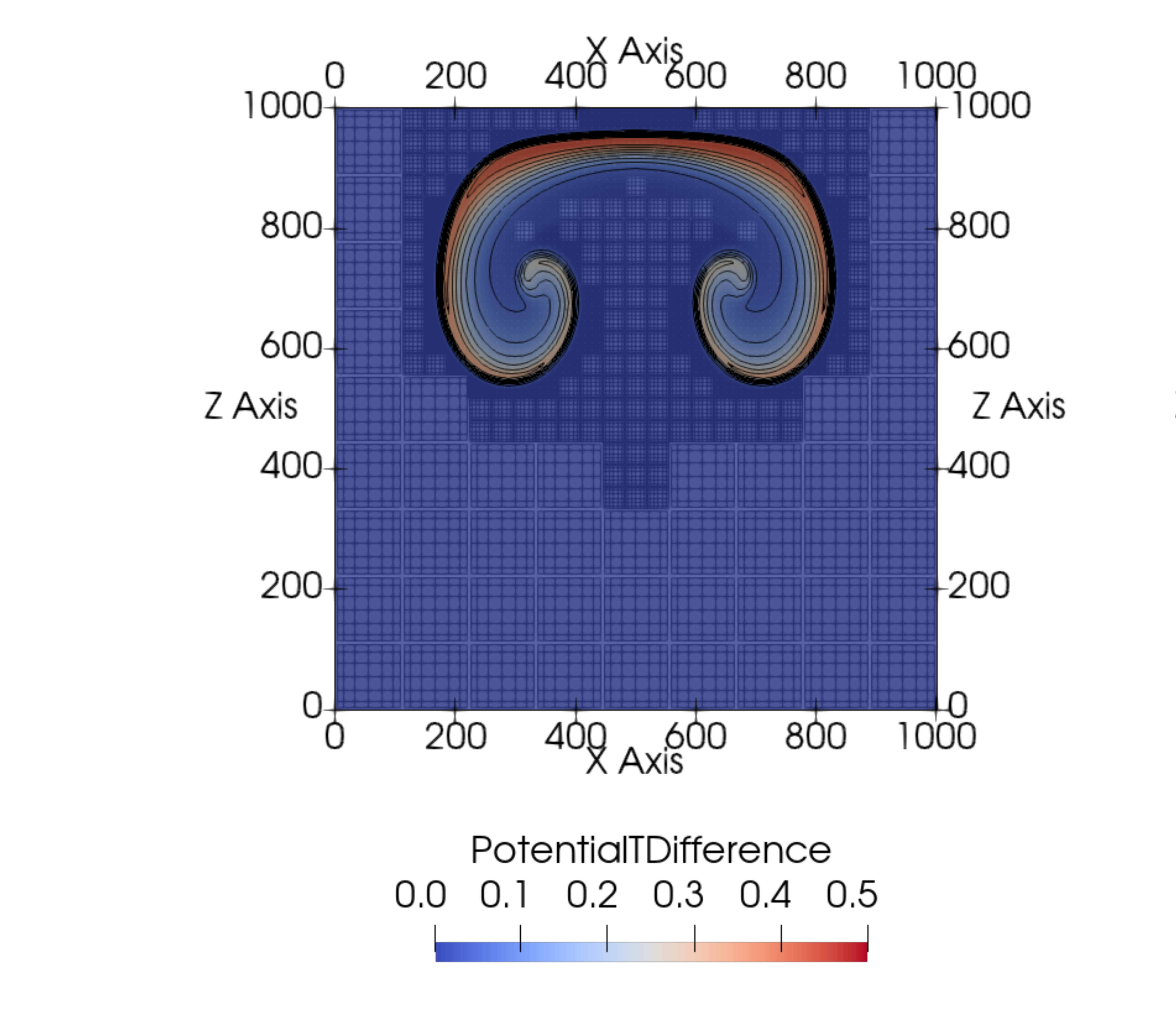}
    \caption{\label{fig:cosine-2d}2D Cosine Bubble.}%
  \end{subfigure}~%
  \begin{subfigure}[t]{0.5\textwidth}
    \centering
    \includegraphics[scale=0.16, trim={42cm 0 0cm 0},clip]{CP032_fig7}
    \caption{\label{fig:cosine-3d}3D Cosine Bubble}
  \end{subfigure}
  \caption{\label{fig:cosine-bubbles-results}Cosine Bubble Scenario.}
\end{figure}

For our second scenario, the cosine bubble, we use a perturbation of the form
\begin{align}
  \label{eq:cos-pertubation}
  \pertubationPotT &= \begin{cases}
    \nicefrac{A}{2} \left[ 1 + \cos(\pi r) \right] & r \leq a, \\
    0 & r > a,
    \end{cases}
\end{align}
where $A$ denotes the maximal perturbation and $a$ is the size of the bubble.
We use the constants
\begin{equation}\label{eq:cosine-bubble}
  A = \SI{0.5}{\K}, \quad a = \SI{250}{\m}, \quad x_c = \SI{500}{\m}, \quad z_c = \SI{350}{\m}.
\end{equation}
For the three-dimensional bubble, we set $y_c = x_c = \SI{500}{\m}$.
This corresponds to the parameters used in~\cite{kelly2012continuous}\footnote{%
We found that the parameters presented in the manuscript of~\cite{kelly2012continuous} only agree with the results, if we use the same parameters as for 2D simulations.}.
For the 2D case, we use a constant viscosity of $\mu = 0.001$ and an \aderdg{}-method of order 6 with two levels of dynamic \amr{}, resulting again in cell sizes of roughly $\SI{111.1}{\m}, \SI{37.04}{\m}, \SI{12.34}{\m}$.
We use slightly different \amr{} parameters of $\Trefine = 1.5$ and $\Tdelete = -0.5$ and let the simulation run for \SI{700}{\s}.
Note that, as seen in \cref{fig:cosine-2d}, our \amr{}-criterion tracks the wavefront of the cloud accurately.
This result shows an excellent agreement to the ones achieved in~\cite{giraldo2008study,muller2010adaptive}.

For the 3D case, we use an \aderdg{}-scheme of order 3 with a static mesh with cell sizes of \SI{40}{\m} and a shorter simulation duration of \SI{400}{\s}.
Due to the relatively coarse resolution and the hence increased aliasing errors, we need to increase the viscosity to $\mu = 0.005$.
This corresponds to a larger amount of smoothing.
Our results (\cref{fig:cosine-3d}) capture the dynamics of the scenario well and agree with the reference solution of~\cite{kelly2012continuous}.

\subsection{Scalability}%
\newcommand{\mdofs}{\text{MDOF/s}}%
All two-dimensional scenarios presented in this paper can be run on a single workstation in less than two days.
Parallel scalability was thus not the primary goal of this paper.
Nevertheless, our implementation allows us to scale to small to medium scale setups using a combined \mpi{}~+~Thread building blocks~(\tbb{}) parallelization strategy, which works as follows:
We typically first choose a number of \mpi{} ranks that ensure an equal load balancing.
\exahype{} achieves best scalability for $1, 10, 83, \ldots$ ranks, as our underlying framework uses three-way splittings for each level and per dimension and an additional communication rank per level.
For the desired number of compute nodes, we then determine the number of \tbb{} threads per rank to match the number of total available cores.

We ran the two bubble scenario for a uniform grid with a mesh size of $729 \times 729$ with order 6, resulting in roughly 104 million degrees of freedom (\textsc{dof}), for 20 timesteps and for multiple combinations of \mpi{} ranks and \tbb{} threads.
This simulation was performed on the SuperMUC-NG system using computational kernels that are optimized for its Skylake architecture.
Using a single \mpi{} rank, we get roughly $4.9$ millions \textsc{dof} updates ($\mdofs$) using two \tbb{} threads and $20.2 \mdofs$ using 24 threads (i.e.\ a half node).
For a full node with 48 threads, we get a performance of $12 \mdofs$.
When using 5 nodes with 10 \mpi{} ranks, we achieve $29.3 \mdofs$ for two threads and $137.3 \mdofs$ for 24 threads.

We further note that for our scenarios weak scaling is more important than strong scaling, as we currently cover only a small area containing a single cloud, where in practical applications one would like to simulate more complex scenarios.
\section{Conclusion}
We presented an implementation of a \muscl{}-scheme and an \aderdg{}-method with \amr{} for the Navier-Stokes equations, based on the \exahypeengine.
Our implementation is capable of simulating different scenarios:
We show that our method has high order convergence and we successfully evaluated our method for standard \textsc{cfd} scenarios:
We have competitive results for both two-dimensional scenarios (Taylor-Green vortex and lid-driven cavity) and for the three-dimensional \textsc{abc}-flow.

Furthermore, our method allows us to simulate flows in hydrostatic equilibrium correctly, as our results for the cosine and colliding bubble scenarios showed.
We showed that our \amr{}-criterion is able to vastly reduce the number of grid cells while preserving the quality of the results.

Future work should be directed towards improving the scalability.
With an improved \amr{} scaling and some fine tuning of the parallelization strategy, the numerical method presented here might be a good candidate for the simulation of small scale convection processes that lead to cloud formation processes.

\subsection*{Acknowledgments}
This work was funded by the European Union’s Horizon 2020 Research and Innovation Programme under grant agreements 
No~671698 (project ExaHyPE, \url{www.exahype.eu}) and 
No~823844 (ChEESE centre of excellence, \url{www.cheese-coe.eu}).
Computing resources were provided by the Leibniz Supercomputing Centre (project pr83no).
Special thanks go to Dominic E.\ Charrier for his support with the implementation in the \exahypeengine{}.

\bibliographystyle{CP032_spmpsci}
\bibliography{CP032_bibliography}{}
\end{document}